\documentclass[final,5p,times,twocolumn]{elsarticle}
\usepackage{color}
\usepackage{amsmath}
\usepackage{amssymb}
\usepackage{subfig}
\usepackage{graphicx}
\usepackage[colorlinks=true,linkcolor=black, citecolor=blue, urlcolor=blue]{hyperref}
\usepackage{lineno,hyperref}
\biboptions{numbers,sort&compress}   % Agrupa las referencias continuas en el texto
\modulolinenumbers[5]

\begin{document}

\begin{frontmatter}

\title{Skyrmions in coupled spin torque nano-oscillator structures}

\author{H. Vigo-Cotrina}
\address{Centro Brasileiro de Pesquisas F\'{\i}sicas, 22290-180,  Rio de Janeiro, RJ, Brazil}

\author{A.P. Guimar\~aes}
\address{Centro Brasileiro de Pesquisas F\'{\i}sicas, 22290-180,  Rio de Janeiro, RJ, Brazil}

%%\author[mymainaddress,mysecondaryaddress]{Elsevier Inc}
%%\ead[url]{www.elsevier.com}

%%\author[mysecondaryaddress]{Global Customer Service\corref{mycorrespondingauthor}}
%%\cortext[mycorrespondingauthor]{Corresponding author}
%%\ead{support@elsevier.com}

%%\address[mymainaddress]{1600 John F Kennedy Boulevard, Philadelphia}
%%\address[mysecondaryaddress]{360 Park Avenue South, New York}

\begin{abstract}
In the present work, using micromagnetic simulation, we show that the magnetic coupling effect plays a very important role in the process of creation of skyrmions in a coupled system of spin-torque nano-oscillators (STNO). First, we have determined the  magnetic ground state in an isolated STNO for different values of perpendicular uniaxial anisotropy (PUA) and Dzyaloshinskii-Moriya interaction (DMI). Next, we have applied a perpendicular pulse polarized current density (J)  and found that it is possible to create a metastable N\'eel skyrmion from a disk whose ground state is a single magnetic domain. From these results, we obtained a phase diagram of polarized current intensity vs. time of application of the current
pulse, for different values of parameters such as PUA, DMI, and distance between the STNOs. Our results show that, depending on the separation distance between the STNOs, the current density required to
create a skyrmion changes due to the magnetic interaction.

\end{abstract}


\end{frontmatter}

%\linenumbers

\section{Introduction}\label{introduction}

Magnetic skyrmions are non-trivial spin textures that can appear in ferromagnets (FM)  where there is lacking inversion symmetry \cite{Shinichiro2016,Guimaraes2017,Sampaio2013,Giovanni2016}. In ultrathin film multilayer systems (FM/substrate), the Dzyaloshinskii-Moriya interaction (DMI) is the main responsible for the creation of skyrmions \cite{Shinichiro2016,Guimaraes2017,Shujun2016}. This DMI has its origin in the interface of the multilayer due to strong spin-orbit coupling between FM and the substrate \cite{Shinichiro2016,Giovanni2016}.\\
There are two skyrmion types: Bloch skyrmion and N\'eel skyrmion
 \cite{Shinichiro2016,Guimaraes2017,Giovanni2016}. The first is generally encountered in bulk systems, while the latter is present in multilayer systems \cite{Sampaio2013,Giovanni2016}. Mathematically, a skyrmion is quantified by  The topological Charge Q, which is defined by Q = (1/4$\pi$)$\int$\textbf{m}.($\partial_x$\textbf{m} $\times
$ $\partial_y$\textbf{m})dxdy, where \textbf{m} is the reduced magnetization. Q = $\pm$1 for skyrmions \cite{Shinichiro2016,Yamane2016}.\\
\indent Several works have shown that a skyrmion  can be stabilized in geometries as, for example, disks \cite{Sampaio2013,Novak2017,Guslienko2018}, ultrathin films \cite{Kim2018} and racetracks \cite{Zhang2015,Tomasello2014}. However, it is also possible to create a multiskyrmion cluster in a single nanodisk \cite{Zhand2015,Liu2017}. Skyrmions can also be created with the help of external perturbations, such as magnetic field\cite{Valle2015,Flovik2017,Masahito2017}, spin  polarized current \cite{Sampaio2013,Lin2013,Yuan2016}, or local heating \cite{Koshibae2014}.\\
\indent Due to the fact that skyrmions are topologically protected  stable magnetic structures \cite{Zhang2016,Guimaraes2017,Sampaio2013} and can be moved with small current densities of the order of  $10^6$ $\,$A/m$^2$ \cite{Jonietz2010}, these have many potential applications, e.g., as logic gate devices \cite{Zhang2015}, racetrack memories \cite{Tomasello2014,Muller2017}, or spin-torque nano-oscillators (STNOs) \cite{Garcia2016,Senfu2015}.\\ 
\indent STNOs with magnetic skyrmions have been object of many studies \cite{Garcia2016,Chui2015,Zhand2015}, since these systems can be used to produce microwave \cite{Kiselev2003} or spin wave based computing and logic devices \cite{Chui2015}.\\
\indent For technological applications, the nanostructures are normally organized in arrays. This leads to the question of magnetic interactions between them. There are in the literature several works about the creation and stabilization of skyrmions in isolated structures \cite{Sampaio2013,Garcia2016,Senfu2015,Yuan2016,Vidal2017,Novak2017,Guslienko2018}, but work is still lacking on the role of magnetic interaction in the process of creation of a skyrmion. The aim of this work is to study the effect of the magnetic coupling in the creation of a N\'eel skyrmion in a spin torque nano-oscillator system. For this purpose, we  tailored the perpendicular uniaxial anisotropy (PUA), Dzyaloshinskii-Moriya interaction (DMI) and polarized current density (J) in the free layer (ultrathin disk) of the STNO (isolated and coupled systems). We used the open source software Mumax3 \cite{Vansteenkiste:2014}, with cell size of 1 $\times$ 1 $\times$ L $\,$nm$^3$, where L is the thickness of the free layer. The material used is Cobalt with parameters \cite{Sampaio2013,Yuan2016,Kim2018}: saturation magnetization M$_s$ = 5.8 $\times$ 10$^5$ A/m, exchange stiffness A = 1.5 $\times$ 10$^{-11}$ J/m, and damping constant $\alpha$ = 0.3 (for faster convergence). The perpendicular uniaxial anisotropy constant (K$_z$) and Dzyaloshinskii-Moriya exchange constant (D$_{int}$) varied from 0.4 to 1.8$\,$MJ/m$^3$ and from 3 to 4.5$\,$mJ/m$^2$, respectively \cite{Sampaio2013,Vidal2017,Novak2017}.  

\section{Results and discussion}
\subsection{Isolated spin torque oscillator}
\begin{figure}[h]
\centering
\includegraphics[width=1\columnwidth]{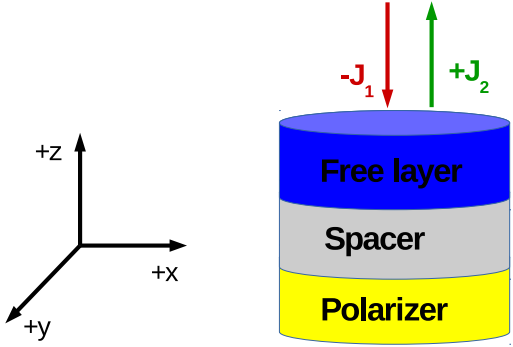}
\caption{Schematic representation of a spin torque nano-oscillator (STNO). Blue region is the free layer, gray region is the spacer and yellow region is the polarizer. J$_1$ and J$_2$ are the densities of the spin currents flowing through the STNO.}\label{oscilador}
\end{figure}

We considered a STNO (see Fig. \ref{oscilador}), where the free layer has thickness L = 0.4$\,$nm and diameter D = 80 $\,$nm. The spacer has parameters: $\Lambda$ = 2 (Slonczewski asymmetry) and $\epsilon$ = 0.2 \cite{Yuan2016}. The polarizer has magnetization \textbf{m}$_{p}$ = (1,0,0)\\

\begin{figure}[h]
\centering
\includegraphics[width=1\columnwidth]{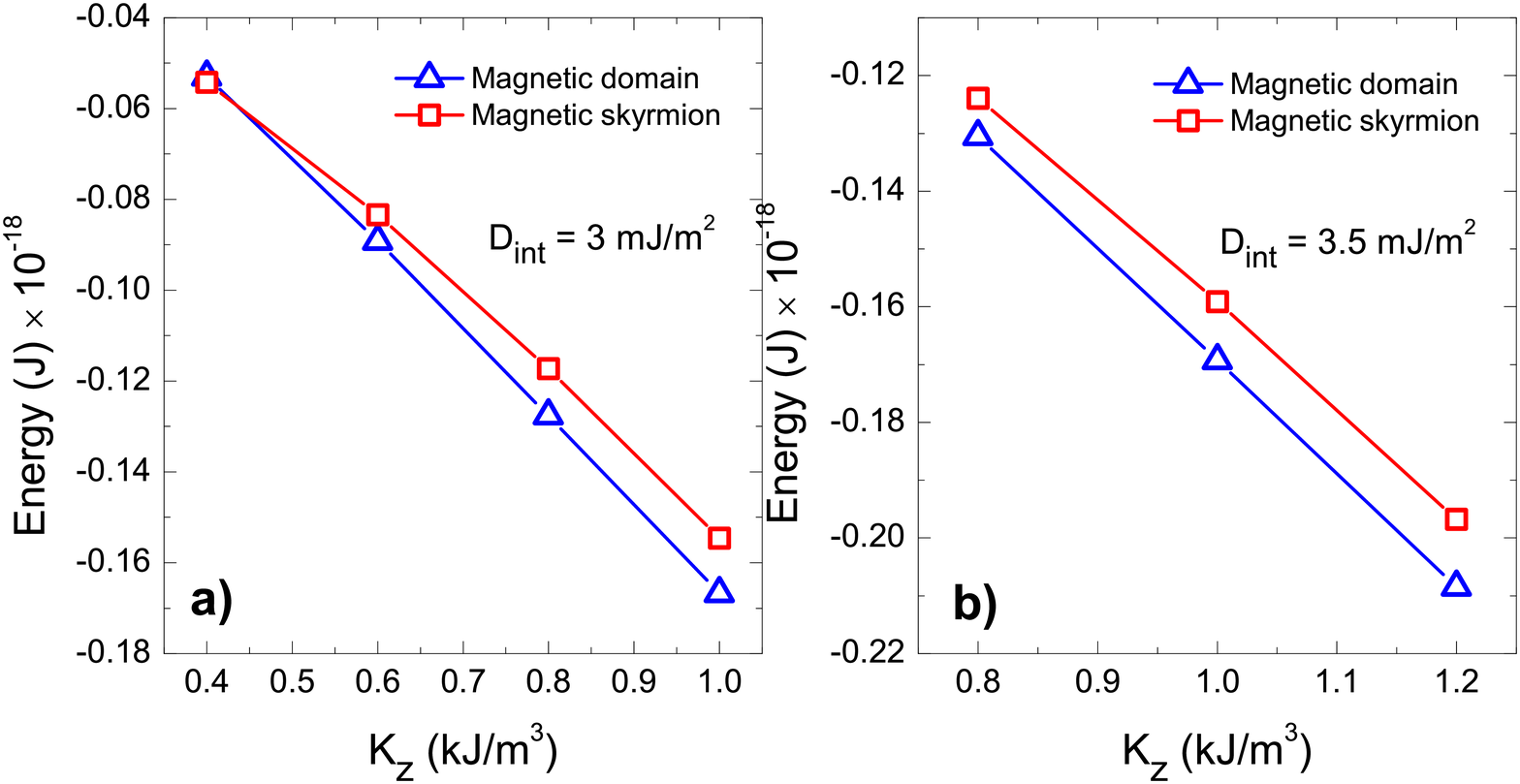}
\includegraphics[width=1\columnwidth]{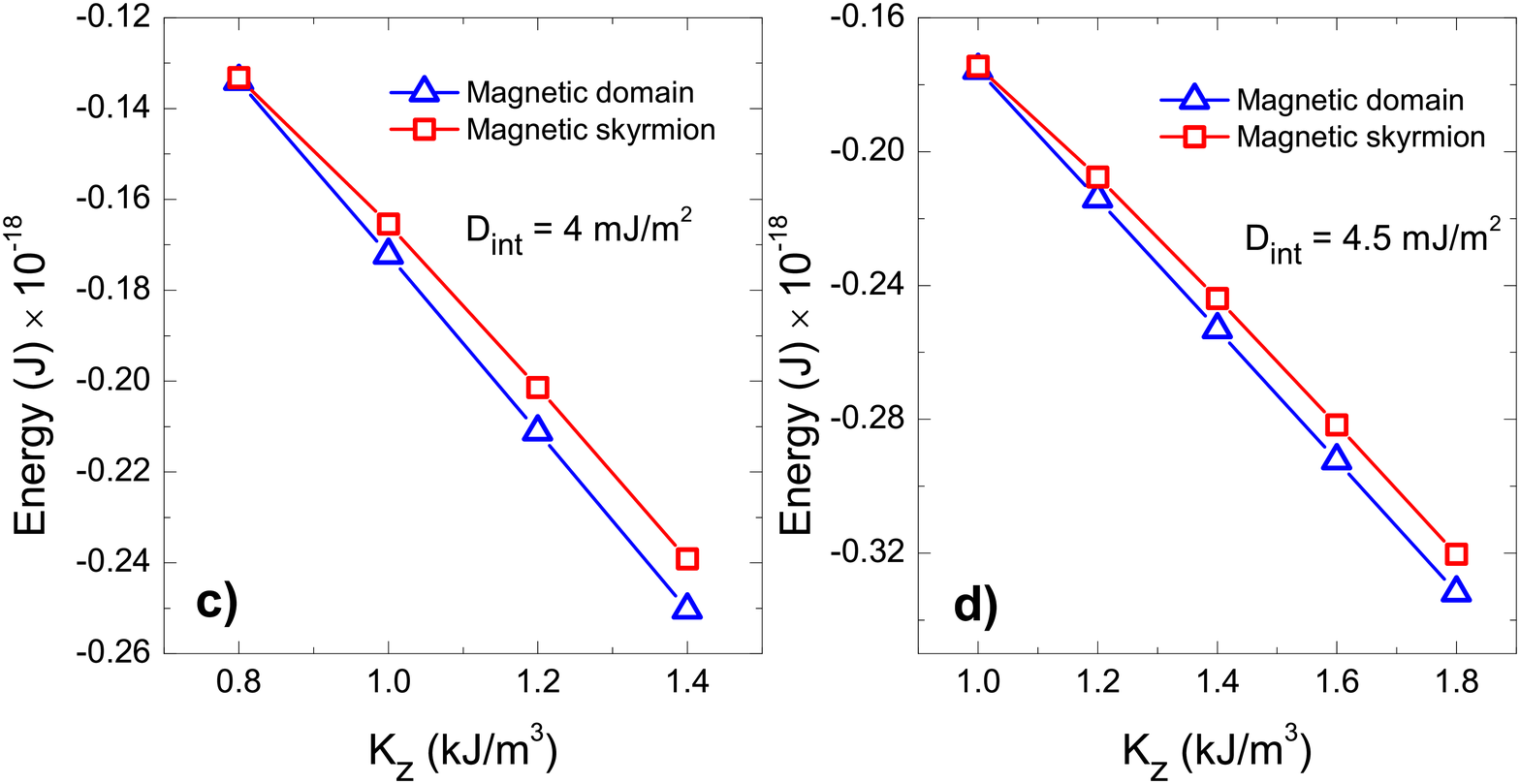}
\caption{Energy of the magnetic final states of the free layer vs. of perpendicular uniaxial anisotropy constant (K$_z$).}\label{energias}
\end{figure}

First, in order to  determine the magnetic ground state of the free layer (J = 0$\,$A/m$^2$), we considered in our micromagnetic simulations two initial magnetic configurations: perpendicular magnetic domain and N\'eel skyrmion. The energies of the final magnetic states are shown in Fig. \ref{energias} for values of DMI, from D$_{int}$ = 3$\,$mJ/m$^2$ to D$_{int}$ = 4.5$\,$mJ/m$^2$. We have considered only cases where both perpendicular magnetic domain (PMD) and magnetic N\'eel skyrmion (MNS) can be obtained as final magnetic states. For example, for D$_{int}$ = 3$\,$mJ/m$^2$ (see Fig. \ref{energias}(a)), the region where it is possible to obtain PMD and MNS as final magnetic states is between K$_z$ = 0.4$\,$MJ/m$^3$ and 
K$_z$ = 1$\,$MJ/m$^3$. For values  less than K$_z$ = 0.4$\,$MJ/m$^3$ (not shown here), there is only one single final magnetic state, that is the magnetic skyrmion, and for values greater that 1$\,$kM/m$^3$ (not shown here), the only final magnetic state is the perpendicular magnetic domain.

\begin{figure}[h]
\centering
\includegraphics[width=1\columnwidth]{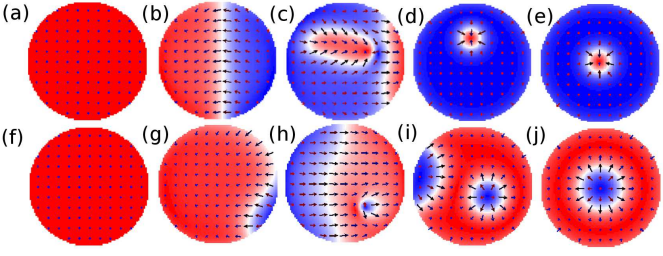}
\caption{Evolution of the z-component of the magnetization configuration from the initial perpendicular single domain to the creation of a metastable N\'eel skyrmion, for D$_{int}$ = 3$\,$mJ/m$^2$ and (a-e)  K$_z$ = 0.8$\,$MJ/m$^3$ and (f-j) K$_z$ = 0.6$\,$MJ/m$^3$}\label{skyrmion1}
\end{figure}

\begin{figure}[h]
\centering
\includegraphics[width=1\columnwidth]{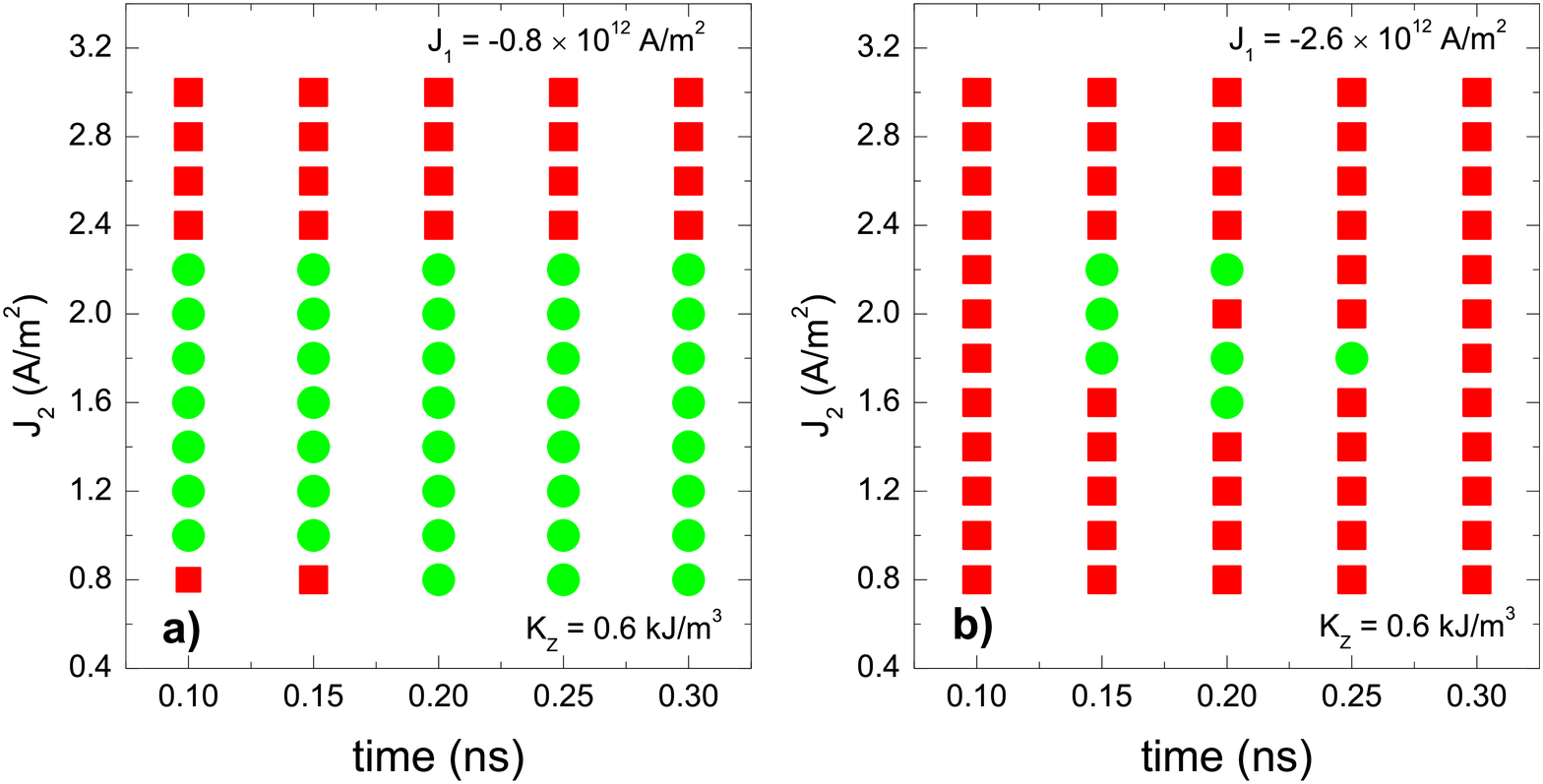}
\includegraphics[width=1\columnwidth]{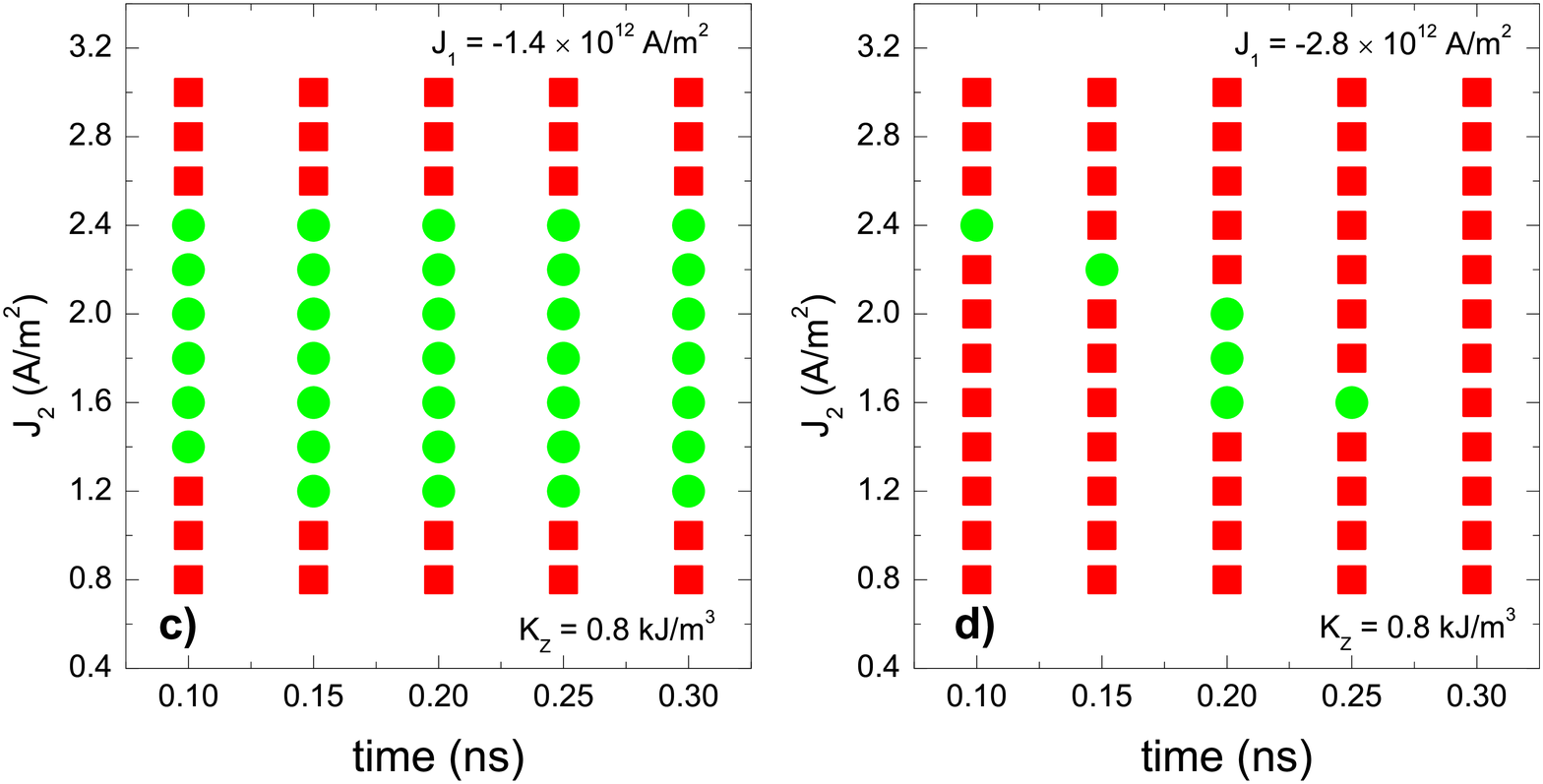}
\caption{Phase diagram for K$_z$ = 0.6$\,$MJ/m$^3$, K$_z$ = 0.8$\,$MJ/m$^3$ and D$_{int}$ = 3$\,$mJ/m$^2$, and different values of the current density J$_1$. Red squares indicate no creation of skyrmion, and green circles indicate creation of skyrmion.}\label{fase1}
\end{figure}

The same behavior occurs for all other cases shown in Fig. \ref{energias}.\footnote{The cases for D$_{int}$ = 2.5$\,$mJ/m$^2$ and D$_{int}$ = 5$\,$mJ/m$^2$ (not shown here) have presented the same behavior.}, for D$_{int}$ = 3.5, 4 and 4.5 mJ/m$^2$.\\
\indent Once established the ranges of values of quantities such as K$_{z}$ and D$_{int}$, we proceed to create a metastable N\'eel skyrmion from the state of lower energy (perpendicular magnetic domain) following the methodology used by Yuan \textit{et al.} \cite{Yuan2016}.\\
\indent There are two currents to be used (see Fig. \ref{oscilador}): -J$_1$ (current flowing in the +z direction) and +J$_2$ (current flowing in the -z direction). We used $\mid$ J$_1$, J$_2$ $\mid$ between 0.8$\,$A/m$^2$ and 3$\,$A/m$^2$. The first was applied  in order to flip the magnetization to the plane of the free layer, in the -x direction, and the last to change the direction to +x. Depending on the value of J$_1$ and the time duration and the value of the current density pulse +J$_2$, it is possible to create a metastable N\'eel skyrmion in the free layer.\\
\indent Unlike the work done by Yuan \textit{et al.} \cite{Yuan2016} where only one case is shown, for $\mid$ J$_1$ $\mid$ = $\mid$ J$_2$ $\mid$ = 2$\times$10$^{12}$A/m$^2$, we here show that considering $\mid$ J$_1$ $\mid$ $\neq$ $\mid$ J$_2$ $\mid$, it is also possible to obtain a MNS. The  topological charge Q was measured in every case to verify if the structure was a skyrmion.\\
\indent First, all the micromagnetic simulations start with perpendicular single domain (Fig. \ref{skyrmion1} (a) and Fig. \ref{skyrmion1} (f)). This magnetic configuration is allowed to relax for a time interval of t = 0.5$\,$ns. After that, in order to flip the magnetization in the -x direction (Fig. \ref{skyrmion1} (b)), -J$_1$ was applied with duration of t = 0.5$\,$ns (fixed time for all simulations). Next, -J$_1$ was switched off,  and the current +J$_2$ was applied to change the direction of magnetization to +x. However, if the duration of J$_2$ is shorter than that required for this purpose, it is possible to see the formation of a deformed N\'eel skyrmion  during this process (Fig. \ref{skyrmion1} (c)). Once obtained this, the system is allowed to relax with J$_2$ = 0, until finally a N\'eel skyrmion is obtained as the final magnetic configuration (\ref{skyrmion1} (e)). For this objective, we have considered duration times for J$_2$ from t = 0.1$\,$ns to 0.3$\,$ns, in 0.05$\,$ns steps. For each value of J$_1$, we used values of J$_2$ from J$_2$ = 0.8$\,$A/m$^2$ to J$_2$ = 3$\,$A/m$^2$ in 0.2$\,$A/m$^2$ steps.\\
\indent The phase diagrams for the cases K$_z$ = 0.6$\,$MJ/m$^3$, K$_z$ = 0.8$\,$MJ/m$^3$  and  D$_{int}$ = 3$\,$mJ/m$^2$ are shown in Fig.\ref{fase1} for different values of J$_1$. It is possible to observe that the phase diagram is also modified by the value of J$_1$, since J$_1$ controls the flipping of the magnetic moments to the plane of the free layer. For example, for K$_z$ = 0.8$\,$MJ/m$^3$ and J$_1$ = -1.4$\times$10$^{12}$A/m$^2$ (Fig. \ref{fase1} (c-d)), we have obtained a metastable N\'eel skyrmion\footnote{The skyrmions  may have polarity p = +1 or p = -1.} for almost the entire range of current duration used here. However, when this value was modified to J$_1$ = -2.8$\times$10$^{12}$A/m$^2$, the phase diagram changes drastically. A similar result is obtained for the case K$_z$ = 0.6$\,$MJ/m$^3$ (Fig. \ref{fase1} (a-b)).

\indent For all values of K$_z$ and D$_{int}$ used here, we have obtained phase diagrams that are dependent on J$_1$, J$_2$ and duration of the current pulse. However, we have not encountered a direct relation between the values J$_1$, J$_2$ and K$_z$, D$_{int}$ in the creation of a metastable skyrmion. The phase diagrams are highly non-linear, nevertheless , it is possible to find some general patterns for the formation of a skyrmion (see Fig. 11 in the Supplementary material\footnote{Supplementary material is abbreviated as SP.}). For example, for the cases where D$_{int}$ = 3.5$\,$mJ/m$^2$, a magnetic skyrmion can be created from a current density threshold\footnote{We focus our attention on J$_1$, since this current density  is the one that is responsible for giving the initial condition to our system.} of $\mid$ J$_1 \mid$ = 1.4$\,$A/m$^2$ for K$_z$ = 0.8$\,$MJ/m$^3$ (Fig. 11 in SP), but when K$_z$ increases to  K$_z$ = 1$\,$MJ/m$^3$, the current density threshold to increase to  $\mid$ J$_1$ $\mid$ = 1.8$\,$A/m$^2$ (Fig. 11 in SP). For K$_z$ = 1.2$\,$MJ/m$^3$, it is not possible anymore  to create a skyrmion with the values of density currents used in this work. The same behavior was found for D$_{int}$ = 4$\,$mJ/m$^2$, the current density threshold increases from $\mid$ J$_1$ $\mid$ = 1.2$\,$A/m$^2$ (for K$_z$ = 0.8$\,$MJ/m$^3$) to $\mid$ J$_1 \mid$ = 1.6$\,$A/m$^2$ (for K$_z$ = 1$\,$MJ/m$^3$) and for K$_z$ = 1.4$\,$MJ/m$^3$, the current density threshold increases up to $\mid$ J$_1$   $\mid$ = 2.8$\,$A/m$^2$. This increase in the values of  current density threshold it is due to the fact that higher values of J$_1$ are necessary in order to flip the magnetization to the plane in order to overcome the perpendicular alignment due to the presence of  K$_z$.\\
\indent On the other hand, it is possible to observe how the increase of D$_{int}$ favors the decrease of the current density threshold $\mid$ J$_1$ $\mid$. For example, for K$_z$ = 0.8$\,$MJ/m$^3$ (Fig. 11 in SP), $\mid$ J$_1$   $\mid$ decreases from $\mid$ J$_1$ $\mid$ = 1.4$\,$A/m$^2$ (D$_{int}$ = 3$\,$mJ/m$^2$) to $\mid$ J$_1$ $\mid$ = 1.2$\,$A/m$^2$ (D$_{int}$ = 4$\,$mJ/m$^2$). The similar behavior is observed for K$_z$ = 1$\,$MJ/m$^3$, where  $\mid$ J$_1$ $\mid$ decreases from $\mid$ J$_1$ $\mid$ = 2.6$\,$A/m$^2$ (D$_{int}$ = 3$\,$mJ/m$^2$) to  $\mid$ J$_1$ $\mid$ = 1.4$\,$A/m$^2$ (D$_{int}$ = 4.5$\,$mJ/m$^2$).\\
\indent In all cases, the skyrmions have topological charge Q $\approx$ 1. Q is not exactly one, because our disk have  finite size.\\ 
\indent An interesting case occur for K$_z$ = 0.8$\,$MJ/m$^3$, D$_{int}$ = 3$\,$mJ/m$^2$, J$_1$ = -3$\,$A/m$^2$, J$_1$ = 2.4$\,$A/m$^2$ an duration time of J$_2$ of t = 10$\,$ns: we have obtained two skymions, as final magnetic configuration, in the free layer, both with polarity p = +1 (in this case the topological charge Q is approximately two).
%\indent In reference \cite{Yuan2016}, the authors argue that in order to create a metastable N\'eel skyrmion, it is necessary  to create with J$_1$ a N\'eel domain wall (Fig. \ref{skyrmion1} (b)), before the application of J$_2$. However, in our simulations we have found that this condition is not necessary. For example,  for K$_z$ = 0.6$\,$kJ/m$^3$ and  D$_{int}$ = 3$\,$mJ/m$^2$, J$_1$ = -0.8$\times$10$^{12}$A/m$^2$  leads to different magnetic configuration (Fig. \ref{skyrmion1} (g)) before applying J$_2$.

\begin{figure}[h]
\centering
\includegraphics[width=1\columnwidth]{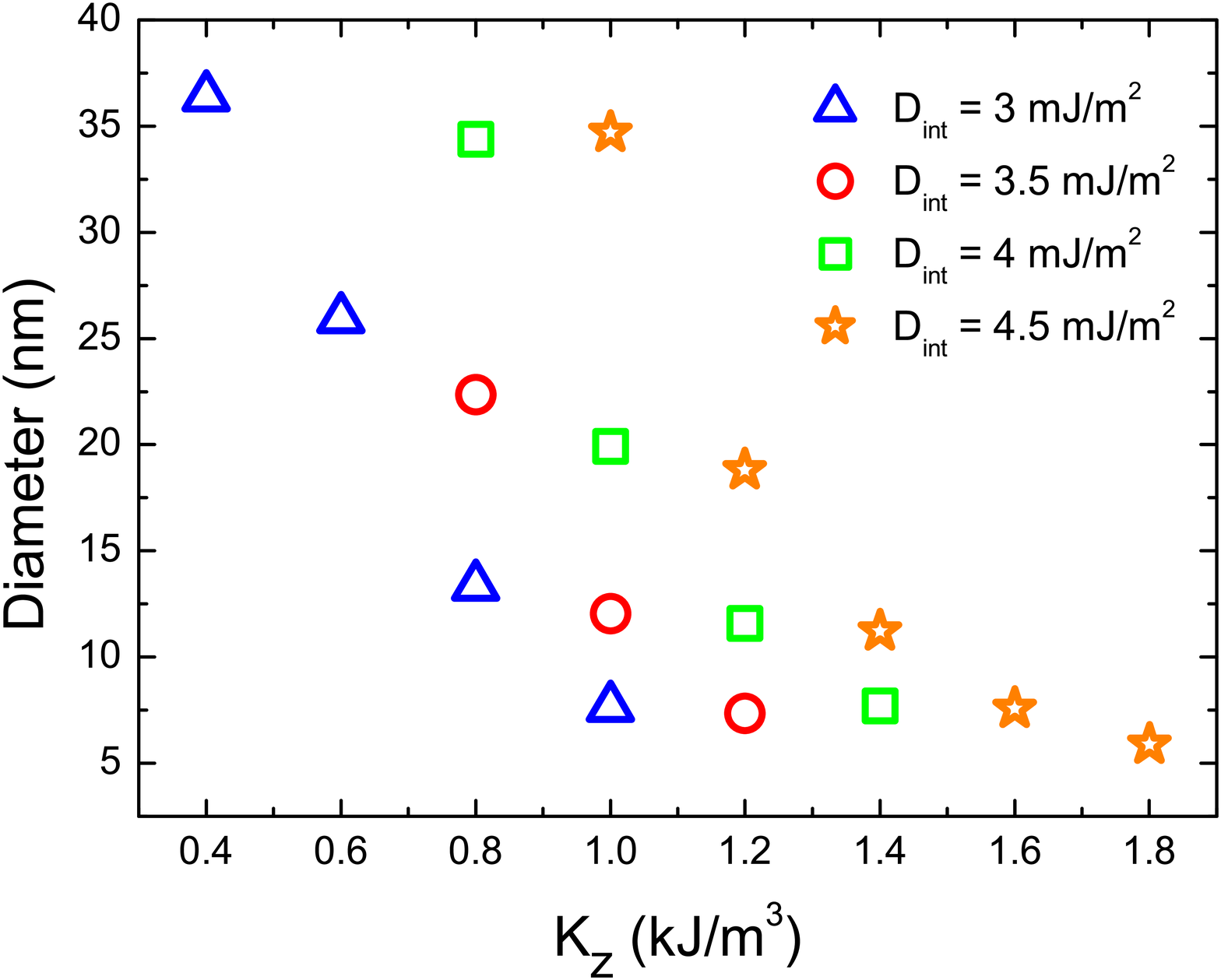}
\caption{Skyrmion diameter D$_N$ as a function of the values of the perpendicular anisotropy constant K$_z$, and D$_{int}$.}\label{diametro}
\end{figure}

\indent For some combinations of K$_z$, D$_{int}$, J$_1$ and J$_2$, we have obtained as final states a  strange magnetic configurations (similar to show in Fig. \ref{skyrmion1} (i)) where a skyrmion coexists with a magnetic domain. These cases, can not be considered as a N\'eel Skyrmion. Similar configurations have been obtained in cobalt disks under application of perpendicular magnetic fields, as already showed by Talapatra \textit{et al.} \cite{Talapatra2018}.\\
\indent We have obtained the average N\'eel skyrmion diameters (D$_N$), considering as D$_N$ diameter, the region where the  z-component of the magnetization (m$_z$) is equal to zero. These values were obtained measuring the Full-Width at Half-Maximum of the profile of m$_z$ along the  diameter of the disk, and their values are shown in Fig. \ref{diametro}. The values of the diameters decrease with the increase of the K$_z$. For example, for D$_{int}$ = 3$\,$mJ/m$^3$, we have obtained a decrease of approximately 80\%,  from D$_N$ $\approx$ 36$\,$nm (K$_z$ = 0.4$\,$MJ/m$^3$) to D$_N$ $\approx$ 8$\,$nm (K$_z$ = 1$\,$MJ/m$^3$). The same behavior occurs for the different values of D$_{int}$. However, the values of D$_N$ increase with the values of D$_{int}$ for the same values of K$_z$. For example, for K$_z$ = 1$\,$MJ/m$^3$, D$_N$ increases from D$_N$ $\approx$ 7.7$\,$nm (D$_{int}$ = 3$\,$mJ/m$^3$) to D$_N$ $\approx$ 34.68$\,$nm (D$_{int}$ = 3$\,$mJ/m$^3$). These behaviors are expected to the fact that the increase of K$_z$ favors the perpendicular alignment of the magnetic moments, thus decreasing the region where m$_z$ = 0, whereas the increase of D$_{int}$ favors the tilting of magnetic moments, increasing the region where m$_z$ = 0.\\
\indent The minimum values of D$_N$, for all combinations of D$_{int}$ and K$_z$, tend to a value
in a range between approximately 5$\,$nm and 7$\,$nm (see Fig. \ref{diametro}), whereas the maximum value obtained of D$_N$ was of approximately 36$\,$nm, which occupies approximately 40$\%$ of the diameter of the  disk (free layer).\\
\indent In the cases where two skyrmions were obtained  in the free layer, both skyrmions have the same D$_N$, whose value is the same as in the case of obtaining a single Skyrmion of approximately D$_N$ $\approx$ 13$\,$nm.\\
\indent Although the N\'eel skyrmions are metastable, we can see that the values of D$_N$ follow the same behavior as in the cases where the skyrmion is a ground state magnetic configuration \cite{Rohart2013}.\\
\indent We have found that the values of D$_N$ do not depend on either current densities (J$_1$ and J$_2$) or duration times of pulse currents. These are dependent only on K$_z$ and D$_{int}$. This case contrasts with the case where the polarizer has magnetization \textbf{m}$_{p}$ = (0,0,1) \cite{Novak2017,Yuan2016}.

\section{Coupled oscillators}
\begin{figure}[h]
\centering
\includegraphics[width=1\columnwidth]{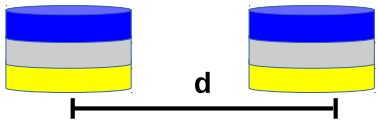}
\caption{Schematic representation of a pair of coupled spin torque oscillators separated by a center to center  distance d.}\label{acoplados}
\end{figure}

\indent In order to study the effect of the magnetic interaction on the creation of skyrmions, we considered a pair of coupled spin torque oscillators (see Fig. \ref{acoplados}).

\begin{figure}[h]
\centering
\includegraphics[width=1\columnwidth]{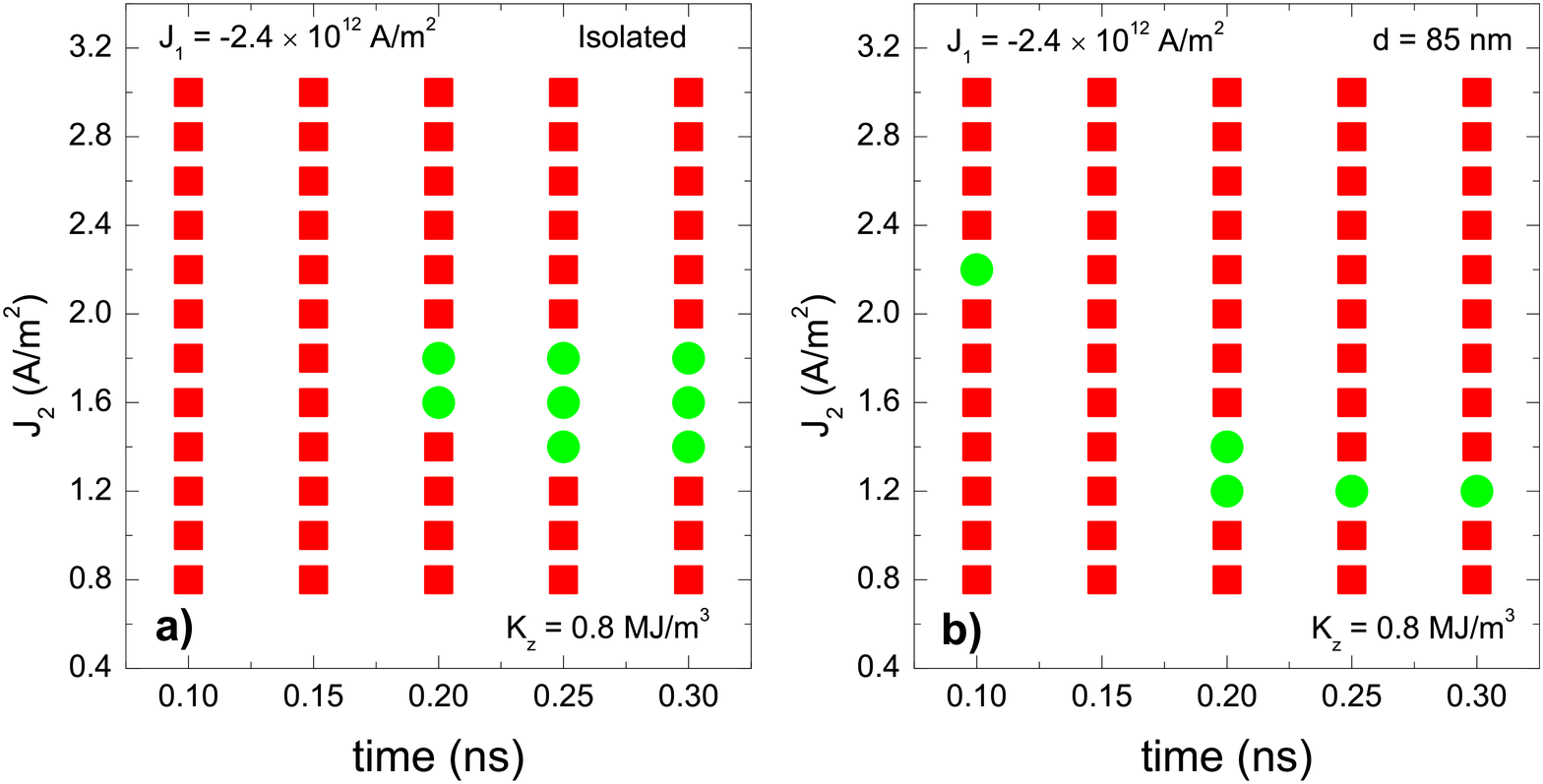}
\includegraphics[width=1\columnwidth]{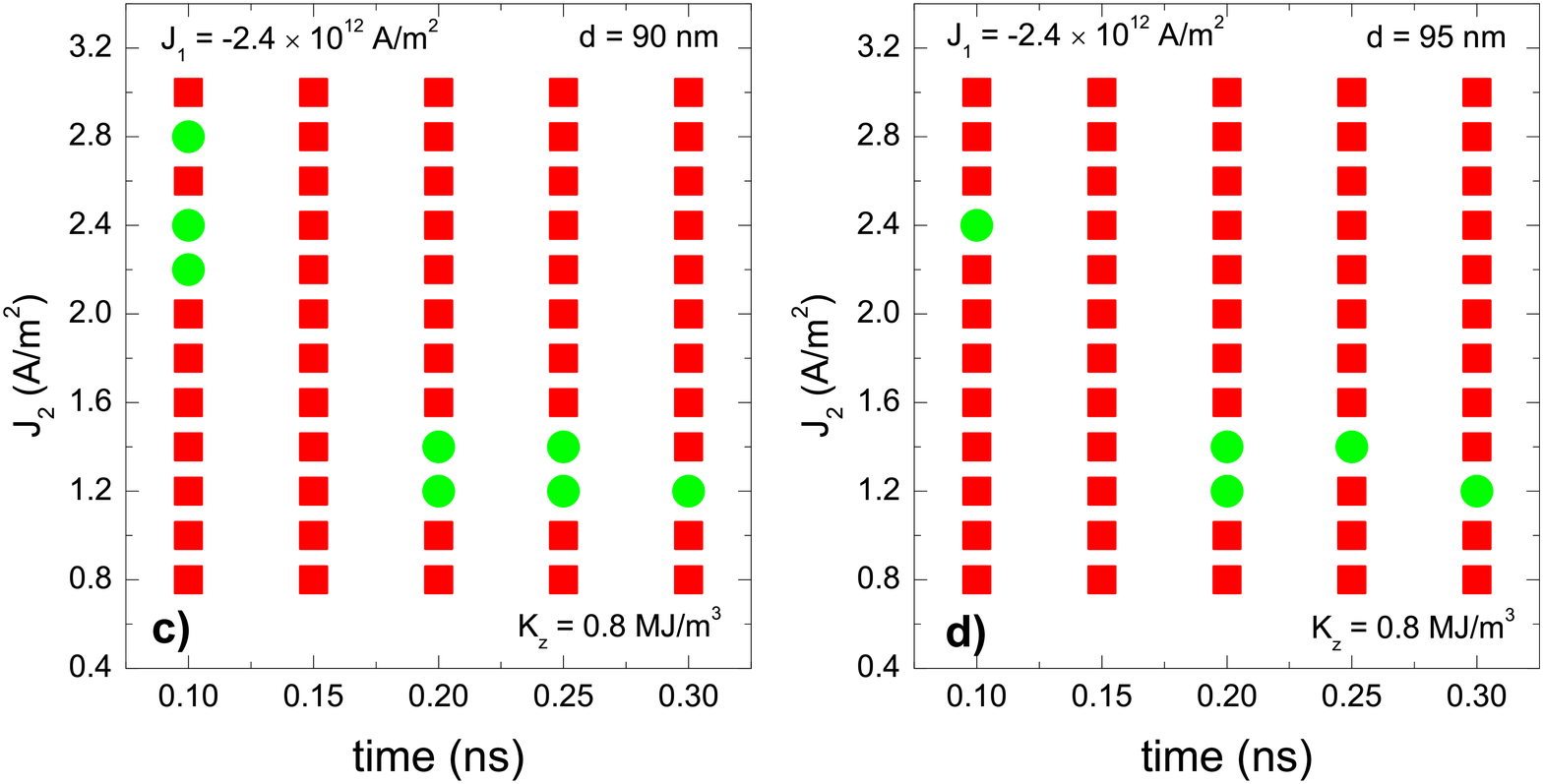}
\caption{Phase diagram for K$_z$ = 0.8$\,$MJ/m$^3$ and D$_{int}$ = 3$\,$mJ/m$^2$ for a) one isolated STNO; b) two STNOs, separation d = 85$\,$nm; c) two STNOs, d = 90$\,$nm; d) two STNOs, d = 95$\,$nm. Red squares indicate no creation of skyrmions in either of the two STNOs, and green circles indicate creation of skyrmions in both STNOs.}\label{acoplados_1}
\end{figure}

\begin{figure}[h]
\centering
\includegraphics[width=1\columnwidth]{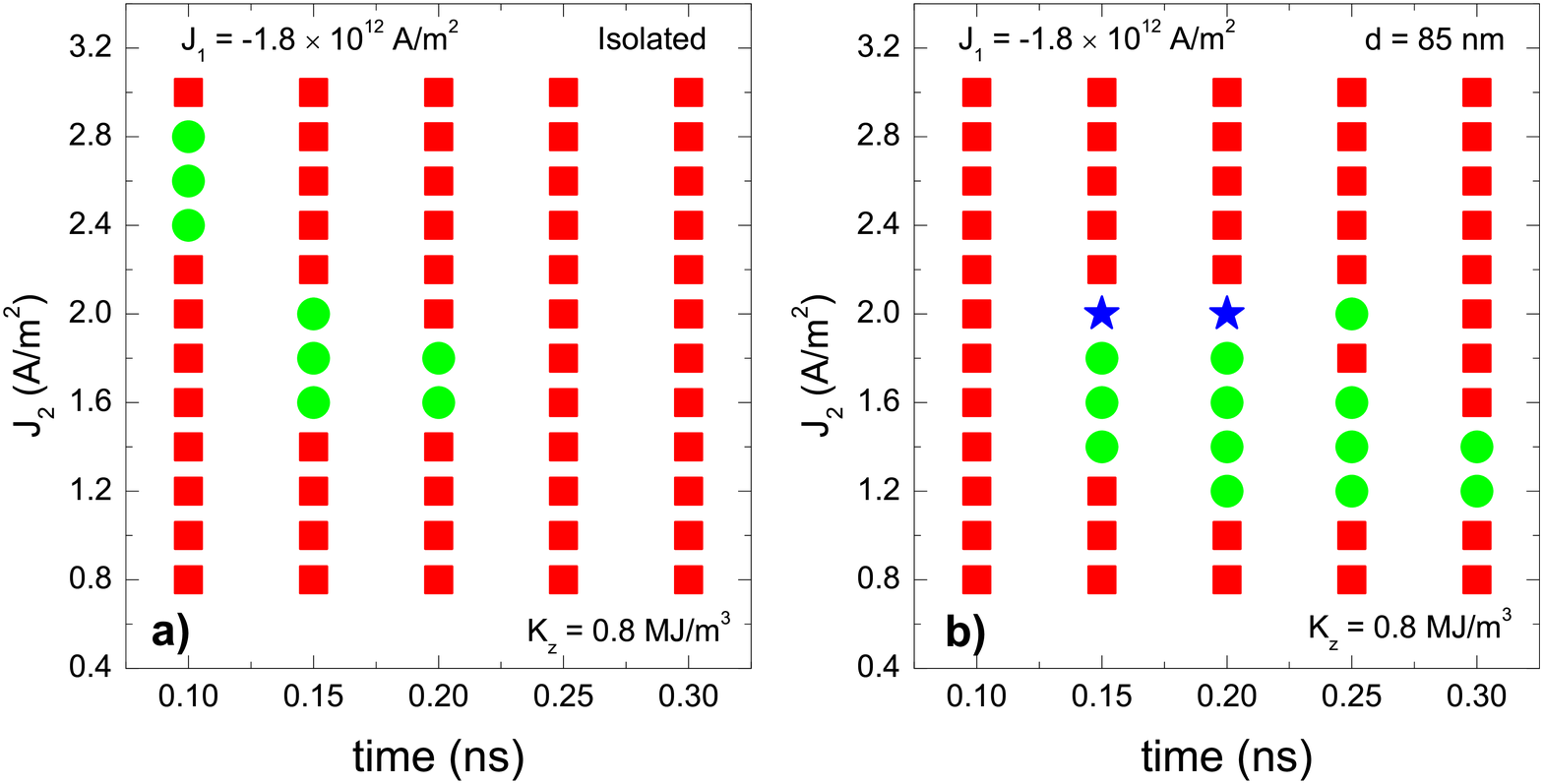}
\includegraphics[width=1\columnwidth]{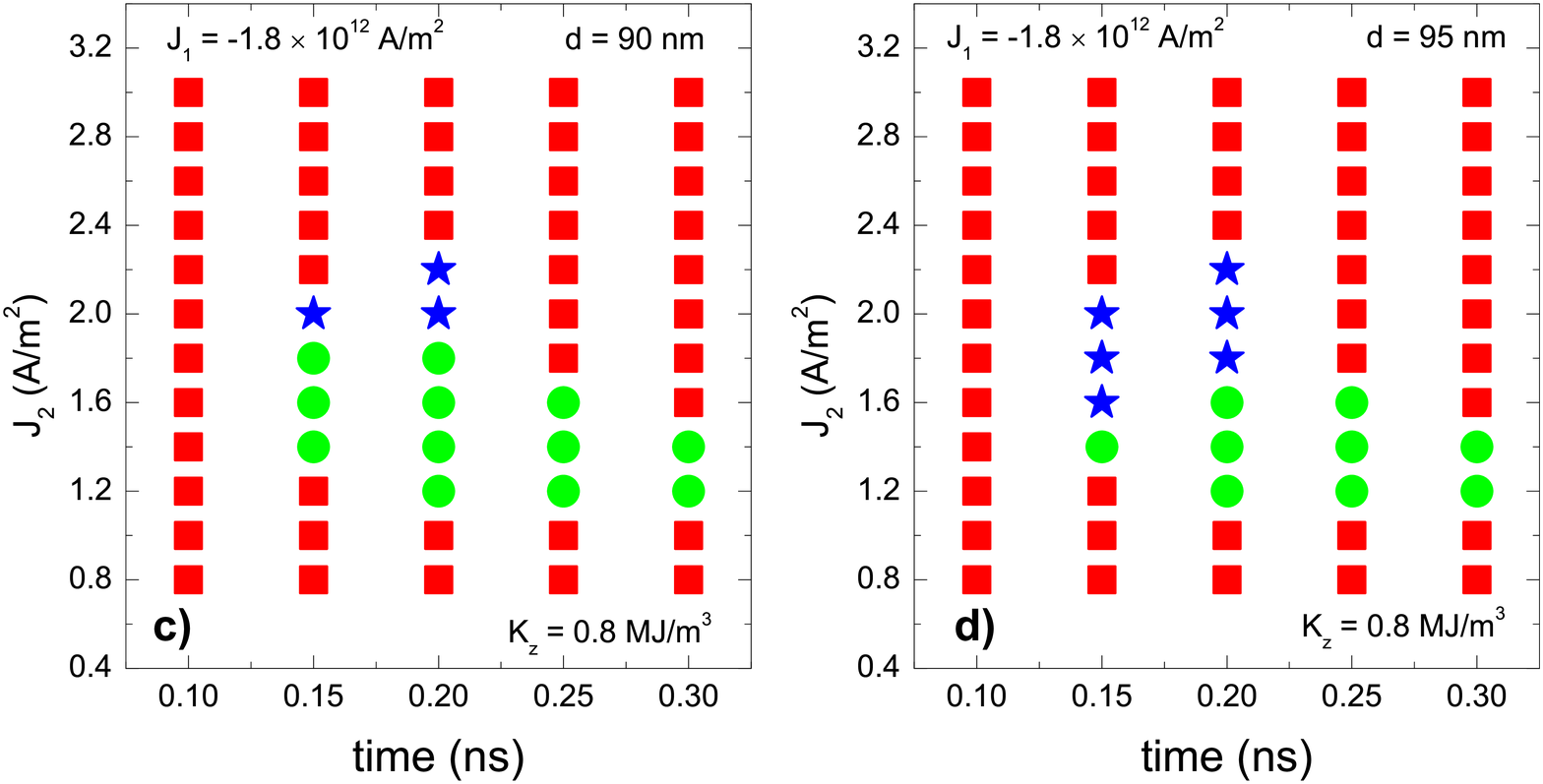}
\caption{Phase diagram for K$_z$ = 0.4$\,$MJ/m$^3$, K$_z$ = 0.8$\,$MJ/m$^3$ and D$_{int}$ = 3$\,$mJ/m$^2$ for a) one isolated STNO; b) two STNOs, separation d = 85$\,$nm; c) two STNOs, d = 90$\,$nm; d) two STNOs, d = 95$\,$nm. Red squares indicate no creation of skyrmion in either of the two STNOs, blue stars indicate the creation of skyrmion in one of the two STNOs, and green circles indicate creation of skyrmions in both STNOs.}\label{acoplados_2}
\end{figure}

We used the same values of K$_z$ and D$_{int}$ shown in Fig. \ref{energias}. The currents J$_1$ and J$_2$ were applied simultaneously in both STNOs in the same way as in the case of an isolated spin torque oscillator (previous section). We have considered a separation center to center  distance d = 85$\,$nm, 90$\,$nm and 95$\,$nm. \\
\indent The new  phase diagrams for the case D$_{int}$ = 3$\,$mJ/m$^2$, K$_z$ = 0.8$\,$MJ/m$^3$ and J$_1$ = -2.4$\times$10$^{12}$A/m$^2$ are shown in Fig. \ref{acoplados_1}. It is possible to observe how the phase diagrams change for different values of the distance d. In some cases the magnetic interaction favors the creation of skyrmions and in other cases it does not. For example, using a value of  J$_2$ = 2.2$\times$10$^{12}$A/m$^2$ and for any value of duration time of J$_2$, it is impossible to create a skyrmion (Fig. \ref{acoplados_1}(a)) in an isolated STNO, but in a coupled system, it is possible to create a skyrmion for separation distances of  d = 5$\,$nm (Fig. \ref{acoplados_1}(b)) and d = 10$\,$nm (Fig. \ref{acoplados_1}(c)), using a duration time of J$_2$ of t = 10$\,$ns.\\
\indent The opposite behavior occurs when  J$_2$ = 1.8$\times$10$^{12}$A/m$^2$. In this case, it is possible to create the skyrmion in an isolated STNO (Fig. \ref{acoplados_1}(a)) for duration time of J$_2$ from t = 20$\,$ns, 25$\,$ns and 30$\,$ns, but it is impossible to create the skyrmion when the STNOs are coupled  (Fig. \ref{acoplados_1}(b), (Fig. \ref{acoplados_1}(c)), (Fig. \ref{acoplados_1}(d))) for any value of duration time of J$_2$.\\
\indent The average N\'eel skyrmion diameters (D$_N$) neither depend on the magnetic interaction nor on the values of  J$_1$ and J$_2$. They have almost the same values as in the case of isolated STNO and have the same dependence on K$_z$ and D$_{int}$ as shown in Fig. \ref{diametro}.\\
\indent The phase diagrams for the coupled systems are also dependent on the values of J$_1$. In Fig. \ref{acoplados_2}, we show the phase diagrams for a value of J$_1$ = -1.8$\times$10$^{12}$A/m$^2$. Its phase diagram is different from  the one obtained for the case J$_2$ = -2.4$\times$10$^{12}$A/m$^2$ (Fig. \ref{acoplados_1}.)

The phase diagrams are shown in Fig. \ref{acoplados_2}. They are more complex than the phase diagrams shown in Fig. \ref{acoplados_1}. Besides the cases where skyrmions were present or absent in both STNOs, there are cases where it was possible to create a skyrmion only in one of the STNOs.  

\begin{figure}[h!]
\centering
\includegraphics[width=1\columnwidth]{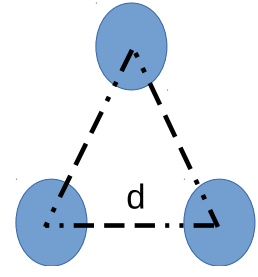}
\caption{Schematic representation of an equilateral triangular array of coupled STNOs separated by a center to center  distance d.}\label{acoplados2}
\end{figure}

\indent The behavior of the current density threshold of $\mid$ J$_1$ $\mid$ (see Fig.12 in SP) follows the same way that in the case of isolated STNO. We can see that the threshold $\mid$ J$_1$ $\mid$ decreases with the increase of D$_{int}$ and $\mid$ J$_1$ $\mid$ increases with the increase of K$_z$. The quantitative values of $\mid$ J$_1$ $\mid$ are almost the same as those in the case of isolated STNO. \\
\indent The results shown in Fig. \ref{acoplados_1} and Fig. \ref{acoplados_2} prove that the magnetic interaction plays an important role in the process of creation of skyrmions in STNOs. In order to further explore the effects of the magnetic interaction, we additionally, have also considered one triangular array\footnote{The procedure to  obtain the skyrmion is the same as in the previous cases.} of STNOs as shown in Fig. \ref{acoplados2}.

\begin{figure}[h!]
\centering
\includegraphics[width=1\columnwidth]{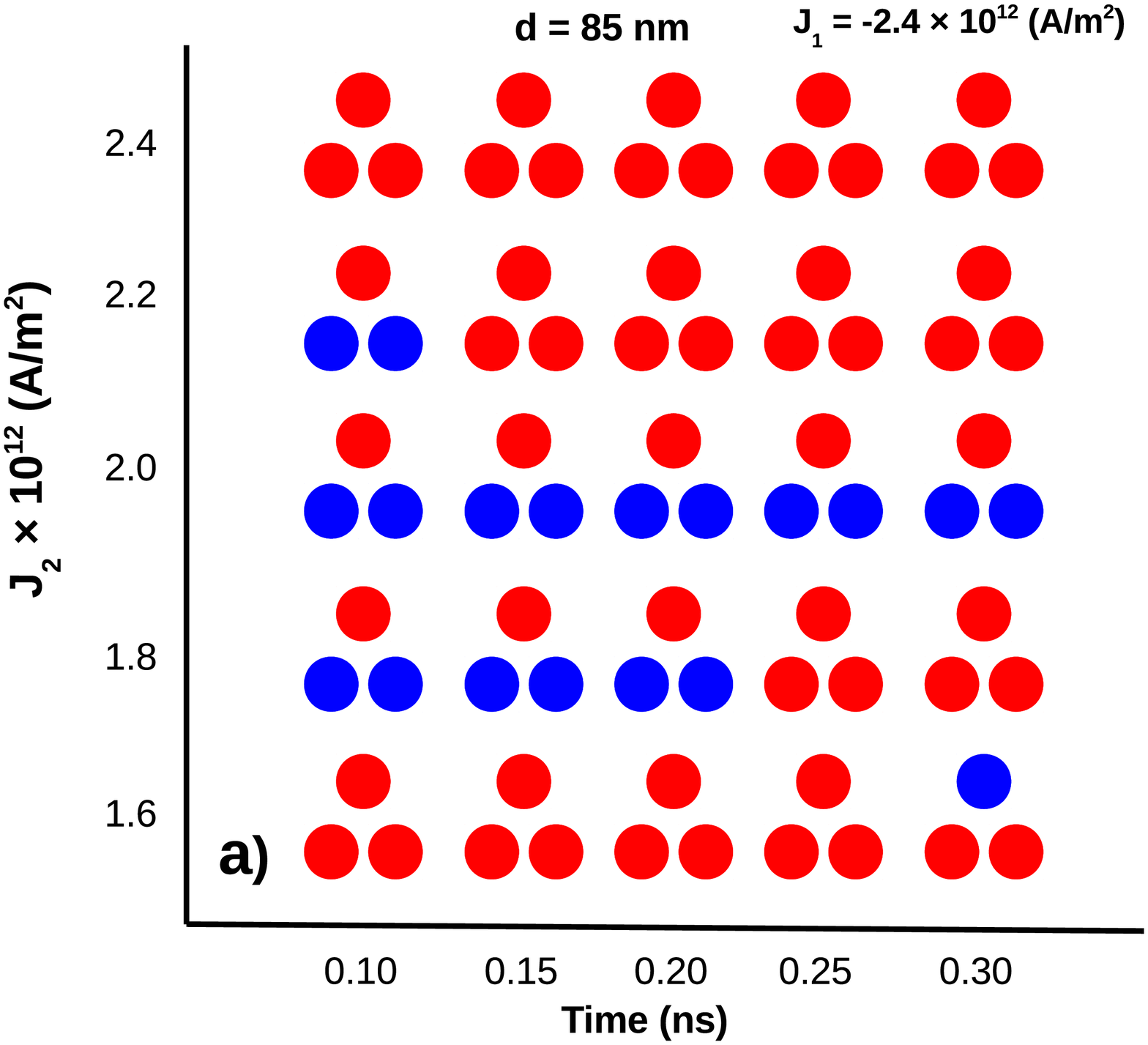}
\includegraphics[width=1\columnwidth]{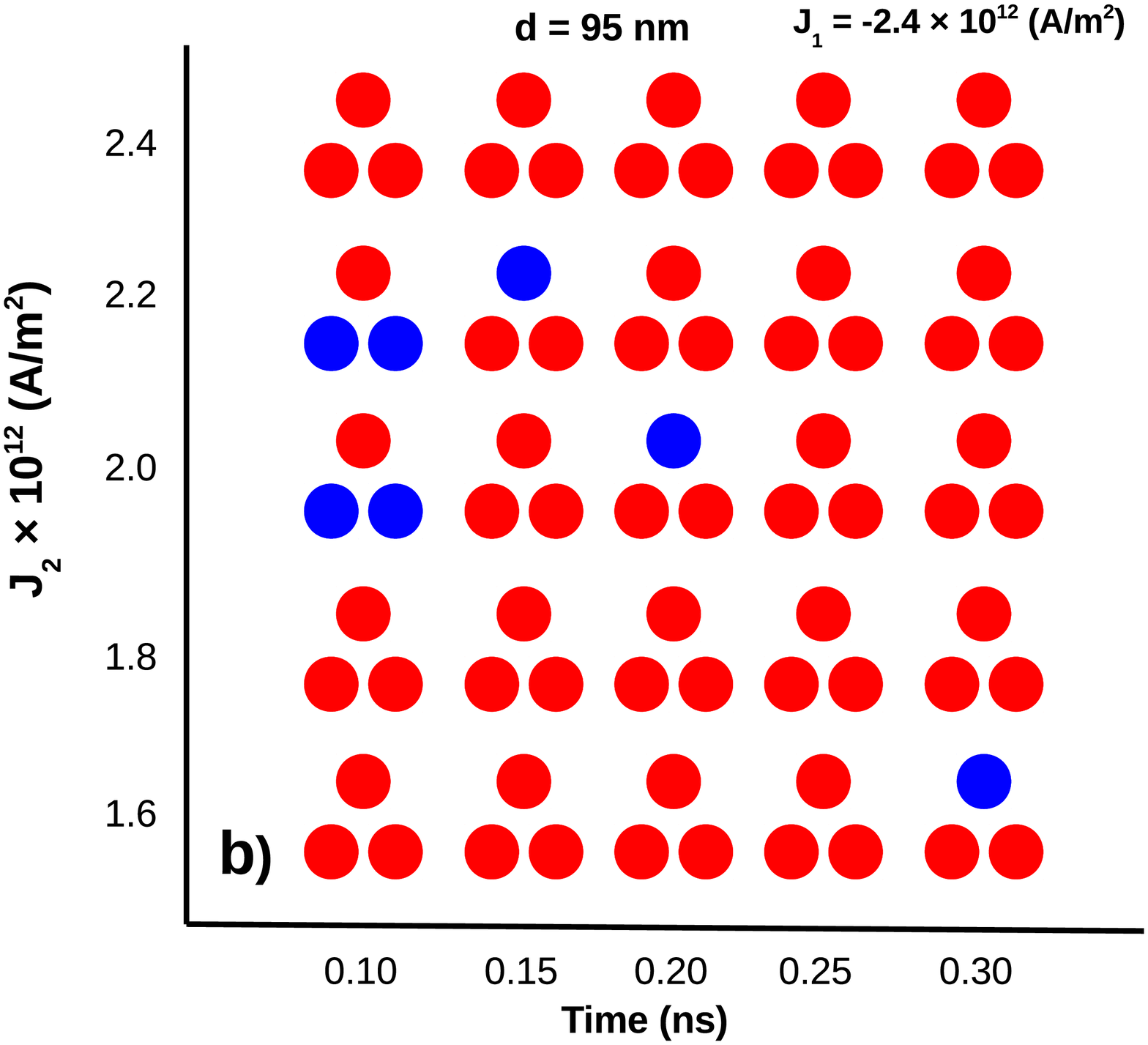}
\caption{Phase diagram, for a triangular array, for K$_z$ = 0.8$\,$MJ/m$^3$ and D$_{int}$ = 3$\,$mJ/m$^2$. Red circles indicate the STNO where there is no creation of skyrmions, green circles indicate the STNO where there is creation of skyrmions}\label{triangulo}
\end{figure}

\indent The phase diagrams for K$_z$ = 0.8$\,$MJ/m$^3$,  D$_{int}$ = 3$\,$mJ/m$^2$, J$_1$ = -2.4$\times$10$^{12}$A/m$^2$ are shown in Fig. \ref{triangulo}. In this figure, it is possible to observe that the creation of skyrmions in the STNOs is different for different distances d.

\indent The STNOs in the triangular array, interacting in a different form from that in the case of a pair of coupled STNOs. For example, for a pair of coupled STNOs, with parameters: J$_2$ = 2.2$\times$10$^{12}$A/m$^2$, t = 0.10$\,$ns and distance d = 85$\,$nm,  the skyrmions can be created in the two STNOs (Fig. \ref{acoplados_1} (a)),  while  for the same parameters, for a triangular array, the  skyrmions cannot be created in the three STNOs (Fig. \ref{triangulo} (a)). Also, a similar behavior can be seen for J$_2$ = 2$\times$10$^{12}$A/m$^2$ and distance d = 85$\,$nm (Fig. \ref{acoplados_1} (b)), while that for a pair of coupled STNOs, the magnetic interaction prevents the creation of skyrmions in the whole  range of values of duration of J$_2$, but this magnetic interaction favors the creation of skyrmions in two of three STNOs (Fig. \ref{triangulo} (a))  

%\indent As expected, the phase diagram for a triangular array also shows dependence on J$_1$ and there is not a simple linear dependence between the parameters: J$_1$, j$_2$ and distance between the STNOs.

\section{Conclusions}
In this work, we have studied the influence of the magnetic interaction in the creation of  a metaestable skyrmion in isolated and coupled systems of STNOs using pulsed spin current densities, using micromagnetic simulation. We have built phase diagrams to know in what conditions it is possible to obtain a skyrmion. Although our  phase diagrams do not show a direct relationship between the parameters used here, we have been able to find some general behavior such as the fact that the threshold J$_1$ increases with the increase of the anisotropy and decreases with the increase of D$_{int}$. Our results show that the  diameters of the metastable skyrmion have the same behavior as that  of stable skyrmions.\\
\indent We have demonstrated that, in the case of coupled systems, the magnetic interaction plays an important role in the creation of metastable skyrmions, modifying the phase diagrams when the distance between the STNOs is changed. However, the magnetic interaction does not modify the values of the threshold J$_1$. \\ 
\indent Our results show a point not sufficiently studied, the influence of the magnetic interaction in the creation of skyrmions.

\section*{ACKNOWLEDGMENTS}
The authors would like to thank the support of the Brazilian agency FAPERJ.
\section*{References}

\bibliography{referencias}

\begin{thebibliography}{10}
\expandafter\ifx\csname url\endcsname\relax
  \def\url#1{\texttt{#1}}\fi
\expandafter\ifx\csname urlprefix\endcsname\relax\def\urlprefix{URL }\fi
\expandafter\ifx\csname href\endcsname\relax
  \def\href#1#2{#2} \def\path#1{#1}\fi

\bibitem{Shinichiro2016}
S.~Seki, M.~Mochizuki, Skyrmions in Magnetic Materials, 1st Edition, Springer,
  Cham, 2016.

\bibitem{Guimaraes2017}
A.~P. Guimar{\~a}es, Principles of Nanomagnetism, 2nd Edition, Springer, Cham,
  2017.

\bibitem{Sampaio2013}
J.~Sampaio, V.~Cros, S.~Rohart, A.~Thiaville, A.~Fert, Nucleation, stability
  and current-induced motion of isolated magnetic skyrmions in nanostructures,
  Nature Nanotechnology 8 (2013) 839–844.
\newblock \href {http://dx.doi.org/10.1038/nnano.2013.210}
  {\path{doi:10.1038/nnano.2013.210}}.

\bibitem{Giovanni2016}
G.~Finocchio, F.~B{\"u}ttner, R.~Tomasello, M.~Carpentieri, M.~Kl{\"a}ui,
  Magnetic skyrmions: from fundamental to applications, Journal of Physics D:
  Applied Physics 49~(42) (2016) 423001.
\newblock \href {http://dx.doi.org/10.1088/0022-3727/49/42/423001}
  {\path{doi:10.1088/0022-3727/49/42/423001}}.

\bibitem{Shujun2016}
S.~Chen, Q.~Zhu, S.~Zhang, C.~Jin, C.~Song, J.~Wang, Q.~Liu, Dynamic response
  for {D}zyaloshinskii–moriya interaction on bubble-like magnetic solitons
  driven by spin-polarized current, Journal of Physics D: Applied Physics
  49~(19) (2016) 195004.
\newblock \href {http://dx.doi.org/10.1088/0022-3727/49/19/195004}
  {\path{doi:10.1088/0022-3727/49/19/195004}}.

\bibitem{Yamane2016}
Y.~Yamane, J.~Sinova, Skyrmion-number dependence of spin-transfer torque on
  magnetic bubbles, Journal of Applied Physics 120~(23) (2016) 233901.
\newblock \href {http://dx.doi.org/10.1063/1.4971868}
  {\path{doi:10.1063/1.4971868}}.

\bibitem{Novak2017}
R.~L. Novak, F.~Garcia, E.~R.~P. Novais, J.~P. Sinnecker, A.~P. Guimar{\~a}es,
  Micromagnetic study of skyrmion stability in confined magnetic structures
  with perpendicular anisotropy, Journal of Magnetism and Magnetic Materials
  451 (2018) 749 -- 760.
\newblock \href {http://dx.doi.org/10.1016/j.jmmm.2017.12.004}
  {\path{doi:10.1016/j.jmmm.2017.12.004}}.

\bibitem{Guslienko2018}
K.~Y. Guslienko, N{\'e}el skyrmion stability in ultrathin circular magnetic
  nanodots, Applied Physics Express 11~(6) (2018) 063007.

\bibitem{Kim2018}
J.~Kim, J.~Yang, Y.-J. Cho, B.~Kim, S.-K. Kim, Coupled breathing modes in
  one-dimensional skyrmion lattices, Journal of Applied Physics 123 (2018)
  053903.
\newblock \href {http://dx.doi.org/10.1063/1.5010948}
  {\path{doi:10.1063/1.5010948}}.

\bibitem{Zhang2015}
X.~Zhang, M.~Ezawa, Y.~Zhou, Magnetic skyrmion logic gates: conversion,
  duplication and merging of skyrmions, Scientific Reports 5 (2015) 9400.
\newblock \href {http://dx.doi.org/10.1038/srep09400}
  {\path{doi:10.1038/srep09400}}.

\bibitem{Tomasello2014}
R.~Tomasello, E.~Martinez, R.~Zivieri, L.~Torres, M.~Carpentieri, G.~Finocchio,
  A strategy for the design of skyrmion racetrack memories, Scientific Reports
  4 (2014) 6784.
\newblock \href {http://dx.doi.org/10.1038/srep06784}
  {\path{doi:10.1038/srep06784}}.

\bibitem{Zhand2015}
S.~Zhang, J.~Wang, Q.~Zheng, Q.~Zhu, X.~Liu, S.~Chen, C.~Jin, Q.~Liu, C.~Jia,
  D.~Xue, Current-induced magnetic skyrmions oscillator, New Journal of Physics
  17~(2) (2015) 023061.
\newblock \href {http://dx.doi.org/10.1088/1367-2630/17/2/023061}
  {\path{doi:10.1088/1367-2630/17/2/023061}}.

\bibitem{Liu2017}
Y.~Liu, H.~Yan, M.~Jia, H.~Du, A.~Du, J.~Zang, Field-driven oscillation and
  rotation of a multiskyrmion cluster in a nanodisk, Phys. Rev. B 95 (2017)
  134442.
\newblock \href {http://dx.doi.org/10.1103/PhysRevB.95.134442}
  {\path{doi:10.1103/PhysRevB.95.134442}}.

\bibitem{Valle2015}
N.~Del-Valle, S.~Agramunt-Puig, A.~Sanchez, C.~Navau, Imprinting skyrmions in
  thin films by ferromagnetic and superconducting templates, Applied Physics
  Letters 107~(13) (2015) 133103.
\newblock \href {http://dx.doi.org/10.1063/1.4932090}
  {\path{doi:10.1063/1.4932090}}.

\bibitem{Flovik2017}
V.~Flovik, A.~Qaiumzadeh, A.~K. Nandy, C.~Heo, T.~Rasing, Generation of single
  skyrmions by picosecond magnetic field pulses, Phys. Rev. B 96 (2017) 140411.
\newblock \href {http://dx.doi.org/10.1103/PhysRevB.96.140411}
  {\path{doi:10.1103/PhysRevB.96.140411}}.

\bibitem{Masahito2017}
M.~Mochizuki, Controlled creation of nanometric skyrmions using external
  magnetic fields, Applied Physics Letters 111~(9) (2017) 092403.
\newblock \href {http://dx.doi.org/10.1063/1.4993855}
  {\path{doi:10.1063/1.4993855}}.

\bibitem{Lin2013}
S.-Z. Lin, C.~Reichhardt, A.~Saxena, Manipulation of skyrmions in nanodisks
  with a current pulse and skyrmion rectifier, Applied Physics Letters 102~(22)
  (2013) 222405.
\newblock \href {http://dx.doi.org/10.1063/1.4809751}
  {\path{doi:10.1063/1.4809751}}.

\bibitem{Yuan2016}
H.~Y. Yuan, X.~R. Wang, Skyrmion creation and manipulation by nano-second
  current pulses, Scientific Reports 6 (2016) 22638.
\newblock \href {http://dx.doi.org/10.1038/srep22638}
  {\path{doi:10.1038/srep22638}}.

\bibitem{Koshibae2014}
W.~Koshibae, N.~Nagaosa, Creation of skyrmions and antiskyrmions by local
  heating, Nature Communications 5~(5148).
\newblock \href {http://dx.doi.org/10.1038/ncomms6148}
  {\path{doi:10.1038/ncomms6148}}.

\bibitem{Zhang2016}
X.~Zhang, Y.~Zhou, M.~Ezawa, High-topological-number magnetic skyrmions and
  topologically protected dissipative structure, Phys. Rev. B 93 (2016) 024415.
\newblock \href {http://dx.doi.org/10.1103/PhysRevB.93.024415}
  {\path{doi:10.1103/PhysRevB.93.024415}}.

\bibitem{Jonietz2010}
F.~Jonietz, S.~M{\"u}hlbauer, C.~Pfleiderer, A.~Neubauer, W.~M{\"u}nzer,
  A.~Bauer, T.~Adams, R.~Georgii, P.~B{\"o}ni, R.~A. Duine, K.~Everschor,
  M.~Garst, A.~Rosch, Spin transfer torques in {M}n{S}i at ultralow current
  densities, Science 330~(6011) (2010) 1648--1651.
\newblock \href {http://dx.doi.org/10.1126/science.1195709}
  {\path{doi:10.1126/science.1195709}}.

\bibitem{Muller2017}
J.~M{\"u}ller, Magnetic skyrmions on a two-lane racetrack, New Journal of
  Physics 19~(2) (2017) 025002.
\newblock \href {http://dx.doi.org/10.1088/1367-2630/aa5b55}
  {\path{doi:10.1088/1367-2630/aa5b55}}.

\bibitem{Garcia2016}
F.~Garcia-Sanchez, J.~Sampaio, N.~Reyren, V.~Cros, J.-V. Kim, A skyrmion-based
  spin-torque nano-oscillator, New Journal of Physics 18~(7) (2016) 075011.
\newblock \href {http://dx.doi.org/10.1088/1367-2630/18/7/075011}
  {\path{doi:10.1088/1367-2630/18/7/075011}}.

\bibitem{Senfu2015}
S.~Zhang, J.~Wang, Q.~Zheng, Q.~Zhu, X.~Liu, S.~Chen, C.~Jin, Q.~Liu, C.~Jia,
  D.~Xue, Current-induced magnetic skyrmions oscillator, New Journal of Physics
  17~(2) (2015) 023061.
\newblock \href {http://dx.doi.org/1367-2630/17/2/023061}
  {\path{doi:1367-2630/17/2/023061}}.

\bibitem{Chui2015}
C.~P. Chui, Y.~Zhou, Skyrmion stability in nanocontact spin-transfer
  oscillators, AIP Advances 5~(9) (2015) 097126.
\newblock \href {http://dx.doi.org/10.1063/1.4930904}
  {\path{doi:10.1063/1.4930904}}.

\bibitem{Kiselev2003}
S.~I. Kiselev, J.~C. Sankey, I.~N. Krivorotov, N.~C. Emley, R.~J. Schoelkopf,
  R.~A. Buhrman, D.~C. Ralph, Microwave oscillations of a nanomagnet driven by
  a spin-polarized current, Nature 425 (2003) 380.
\newblock \href {http://dx.doi.org/10.1038/nature01967}
  {\path{doi:10.1038/nature01967}}.

\bibitem{Vidal2017}
N.~Vidal-Silva, A.~Riveros, J.~Escrig, Stability of {N}{\'e}el skyrmions in
  ultra-thin nanodots considering {D}zyaloshinskii-{M}oriya and dipolar
  interactions, Journal of Magnetism and Magnetic Materials 443 (2017) 116 --
  123.
\newblock \href {http://dx.doi.org/10.1016/j.jmmm.2017.07.049}
  {\path{doi:10.1016/j.jmmm.2017.07.049}}.

\bibitem{Vansteenkiste:2014}
A.~Vansteenkiste, J.~Leliaert, M.~Dvornik, M.~Helsen, F.~Garcia-Sanchez,
  B.~Van~Waeyenberge, The design and verification of {M}umax3, AIP Advances 4
  (2014) 107133.
\newblock \href {http://dx.doi.org/10.1063/1.4899186}
  {\path{doi:10.1063/1.4899186}}.

\bibitem{Talapatra2018}
A.~Talapatra, J.~Mohanty, Scalable magnetic skyrmions in nanostructures,
  Computational Materials Science 154 (2018) 481 -- 487.
\newblock \href {http://dx.doi.org/10.1016/j.commatsci.2018.08.022}
  {\path{doi:10.1016/j.commatsci.2018.08.022}}.

\bibitem{Rohart2013}
S.~Rohart, A.~Thiaville, Skyrmion confinement in ultrathin film nanostructures
  in the presence of {D}zyaloshinskii-{M}oriya interaction, Phys. Rev. B 88
  (2013) 184422.
\newblock \href {http://dx.doi.org/10.1103/PhysRevB.88.184422}
  {\path{doi:10.1103/PhysRevB.88.184422}}.

\end{thebibliography}

\end{document}